# Hydrodynamic effects on biofilms at the bio-interface using a microfluidic electrochemical cell: case study of *Pseudomonas sp.*

*Mirpouyan Zarabadi[1], François Paquet-Mercier[1], Steve J. Charette[2,3,4], Jesse Greener[*,1]*

[1]Département de Chimie, Faculté des sciences et de génie, Université Laval, Québec City, QC, Canada.

[2]Institut de Biologie Intégrative et des Systèmes, Pavillon Charles-Eugène-Marchand, Université Laval, Québec City, QC, Canada.

[3]Centre de recherche de l'Institut universitaire de cardiologie et de pneumologie de Québec, Québec City, QC, Canada.

[4]Département de biochimie, de microbiologie et de bio-informatique, Faculté des sciences et de génie, Université Laval, Québec City, QC, Canada.

**ABSTRACT**

The anchoring biofilm layer is expected to exhibit a different response to environmental stresses than for portions in the bulk, due to the protection from other strata and the proximity to the attachment surface. The effect of hydrodynamic stress on surface-adhered biofilm layers was tested using a specially designed microfluidic bio flow cell with an embedded three-electrode detection system. *In situ* electrochemical impedance spectroscopy (EIS) measurements of biocapacitance and bioresistance of *Pseudomonas sp.* biofilms were conducted during the growth phase and under different shear flow conditions with verification by other surface sensitive techniques. Distinctive, but reversible changes to the amount of biofilm and its structure at the attachment surface were observed during the application of elevated shear stress. In contrast, regular microscopy revealed permanent distortion to the biofilm bulk, in the form of streamers and ripples. Following the application of extreme shear stresses, complete removal of significant portions of biofilm outer layers occurred, but this did not change the measured quantity of biofilm at the electrode attachment surface. The structure of the remaining biofilm, however, appeared to be modified and susceptible to further changes following application of shear stress directly to the unprotected biofilm layers at the attachment surface.

**Introduction**

Biofilm formation enables the firm attachment of sessile bacteria to solid surfaces where they can develop an environmental niche. Tangential (shear) liquid flow can aid the bacteria by enhancing mass transfer of essential molecular species into the biofilm and preventing buildup of molecular byproducts. This has a pronounced impact on biofilm mechanical properties[1] and physical properties,[2-5] but also on the metabolism,[6] growth kinetics,[7,8] and biochemistry[9,10] of the resident microorganisms. The impact of time-changing flow on biofilm properties is less understood.[11,12] The effect of fluidic conditions on the biofilm is complicated by the fact that flow against it is expected to affect outer layers disproportionately, whereas portions at the attachment surface (the bio-interface) remain relatively isolated.[13,14] For example, much of the biofilm resistance to elimination is linked to the surface attached portions, where highly attenuated molecular mass transfer of nutrients and oxygen may give rise to so-called "persister" cells.[15,16] Many biofilm species are also known to have localized anchoring points, resulting in mushroom-like structures.[17] These are critical features that allow limited convection through the interior of the biofilm.[14,16] Biofilm structure and the influence of shear stress on it can be studied by optical microscopy, but the details at the attachment surface are hard to resolve due to interference from other strata. Surface-sensitive techniques can help resolve this problem. Confocal laser scanning microscopy (CLSM) can reject optical interference originating away from the attachment surface. However, even with fluorescent protein expressing bacteria, the extracellular polymeric material remains invisible and attempts to add fluorescent markers does not guarantee uniform results in space or time.[18] While, quantitative studies using high-resolution imaging are difficult to capture global properties, such measurements can add invaluable insights and confirmation related to structural properties. Despite the fact that attachment to highly polished optical surfaces are expected to be different than for other surfaces, attenuated total reflection infrared (ATR-IR) spectroscopy has been successfully used to measure biofilm biomass and extracellular matrix materials within a few microns from the surface under flow conditions. Other surface techniques such as surface plasmon resonance spectroscopy and surface enhance Raman spectroscopy, can be valuable for biofouling studies, but require specific surface preparation.[19,20] Electrochemical techniques have been gaining favour as versatile approaches in studying bacteria attachment and biofilm development.[21-24] These approaches are also gaining attention in the microfluidics community,[25-27] thanks in part to their advances in microfabrication techniques.

Here we present a compact three-electrode system embedded in a microfluidic flow cell for studies of *Pseudomonas sp.* biofilms exposed to laminar flow conditions. Electrochemical impedance spectroscopy (EIS) produced Nyquist plots, which were subsequently analyzed using an accepted equivalence circuit for non-electroactive biofilms. This produced time-changing equivalent biofilm capacitance and resistance values, which helped assess biomass accumulation and structural changes as a function of growth time and hydrodynamic conditions. Parallel microscopy provided comparison with bulk biofilm properties, whereas ATR-IR and CLSM measurements were used to verify EIS observations regarding biofilm accumulation and structure at the bio-interface.

**Experimental**

**Fabrication of a microfluidic flow cell and *in situ* EIS**

The graphite (GraphiteStore.com Inc., USA) working electrode (WE) and counter electrode (CE) were cut into 3 mm × 20 mm strips. A pseudo-reference electrode was made by electroless deposition of gold on a planar polystyrene substrate. The solution used for this included 10 mM of 99.999% chloroauric acid (Sigma-Aldrich, USA) as a gold source, with 18 mM D-glucose (Sigma-Aldrich, USA) as a reducing agent and 400 mM sodium bicarbonate (Sigma-Aldrich, USA) as a pH buffer in distilled water. After degassing, the solution was poured on a planar polystyrene substrate and was kept for 24 h at ambient temperature. The gold coated polystyrene was then rinsed with distilled water and cut into 5×20 mm strips.

Figure 1 shows the device fabrication steps. All three electrodes were placed in direct contact with the photoresist features on a silicon master mould (FlowJEM Inc., Toronto, Canada) using double-sided tape. Next, a 10:1 mixture of liquid polydimethylsiloxane (PDMS) and cross-linking agent Sylgard184 (Dow corning, Canada) was poured over the mould and left to cure for 4 h at 70 °C. This encased the three electrodes in PDMS except for the portion that was protected by the feature surface on the mould. Once cured, the PDMS device was carefully removed from the mould. As PDMS was excluded from flowing between the electrode and the channel feature on the mould, the surface of the electrodes could be exposed at the bottom of the microchannels after removing the double-sided tape. This resulted in a smooth transition from the PDMS portion of the microchannel to the electrodes. A portion of the PDMS device was then removed, exposing the electrodes for connection to the potentiostat. The sequential orientation of WE and CE relative to the flow direction in the microchannel is similar to other devices used for chemical sensing[28] and biofilm studies by voltammetry and EIS.[25-27] Two inlets were punched into the channel. The first ($I_1$) was at the far end, just before the gold reference electrode, and the second ($I_2$) just after the gold electrode (Figure 1d). A third hold was punched at the opposite end of the channel as a device outlet (O). Cleaning and sterilization of channel and electrodes was conducted before sealing. Sterile cotton swabs dipped in 1M HCl were used for cleaning graphite electrodes and 70% ethanol for the PDMS channel surfaces and gold electrode. The time of exposure and applied scrubbing force did not visibly damage to the electrodes or swell of the PDMS, nor were any changes to the electrical conductivity detected after cleaning. A microscope glass slide (Fisher Scientific, USA) was cleaned by immersion in 70% ethanol for 5 minutes and then dried under clean $N_2$ gas stream in a laminar flow cabinet. The microfluidic electrochemical cell and microscope slide were placed into a plasma cleaning oven (PCD-001 Harrick Plasma,

Ithaca, NY, USA) for 90 seconds exposure to sterile air plasma at 650 mtorr. This further sterilized both components and enabled surface activation of the PDMS for irreversible bonding. Sterile perfluoroalkoxy connective tubing (Hamilton Inc., Canada) was attached to inlets, which were in turn connected to 60 mL syringes (BD Scientific, NJ, USA) via connector assemblies (P-200x, P-658, IDEX, WA, USA). Syringe pumps (PHD 2000, Harvard Apparatus, Holliston, MA, USA) were used for liquid injection. The microfluidic channel and tubing were then rinsed with sterile distilled water for 1 hour at 1 mL·h$^{-1}$ through inlets $I_1$ and $I_2$. Finally, electrical connections were made between the electrodes and the potentiostat.

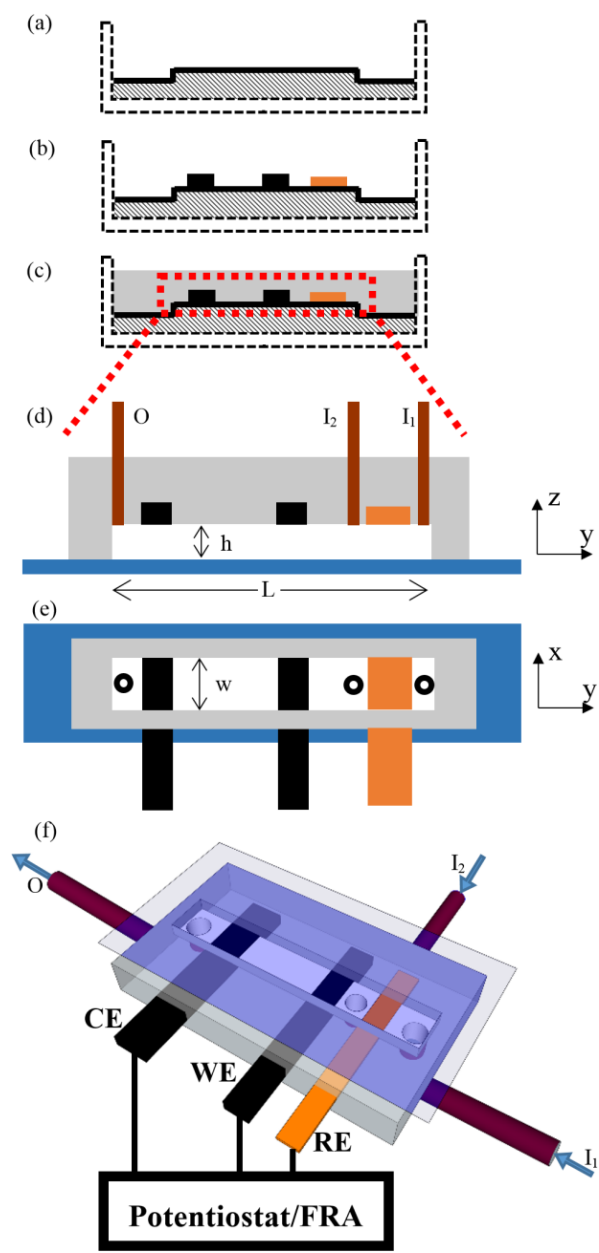

**Figure 1.** Device fabrication. (a) A cross-section view of the mould (diagonal cross-hatch) with a raised feature that defines the microchannel, which is fixed to the bottom of a container (dashed). (b) Two 3 mm wide graphite electrodes (black) and a 5 mm wide gold electrode (yellow) were placed on top of the mould channel feature. (c) PDMS (grey) with cross-linker were poured over the mould/electrode assembly. (d) A zoomed cross-section view from the marked region in (c) of the microfluidic device after removal from the mould with embedded electrodes and sealing to a glass microscope slide (blue). Device includes inlets ($I_1$, $I_2$), and outlet (O). Cleaning process described in text not shown. (e) Bird's eye view of the microfluidic electrochemical cell in (d). (f) Thee-dimensional rendering of the microfluidic device with embedded working (WE), counter (CE) and gold pseudo-reference electrodes (RE) connected to a potentiostat with frequency response analyzer (FRA). Arrows show the direction of flow into inlets $I_1$ and $I_2$ and out of outlet O. Electrode sizes are not to scale. Channel length, height and width were L=33 mm, h=300 μm and w=2 mm, respectively, as defined by the mould protrusion in (a). The axes x and y are perpendicular and parallel to the flow direction of the channel, respectively, and z is parallel with the channel vertical axis.

**Bacteria and solution preparation**

The bacterial strain used was *Pseudomonas fluorescens* CT07 tagged with a chromosomally integrated green fluorescent protein (GFP), known as *Pseudomonas sp.*, strain CT07 in the literature.[29] The nutrient source used in this study was of the AB type, which consisted of 1.51 mM $(NH_4)_2SO_4$, 3.37 mM $Na_2HPO_4$, 2.20 mM $KH_2PO_4$, 179 mM NaCl, 0.1 mM $MgCl_2·6H_2O$, 0.01 mM $CaCl_2·2H_2O$, 0.001 mM $FeCl_3$ and 10 mM $Na_3C_6H_5O_7$ (trisodium citrate) as the carbon source. Liquid cultures were obtained by transferring and dispersing bacterial colonies from an agar-culture dish containing AB growth medium and to 3 mL of liquid AB growth medium. The pre-culture was ready for use after 18 h of incubation at 30 °C on an orbital shaker operating at 300 rpm. This protocol was followed precisely to obtain inoculum solutions with similar bacterial activities from experiment to experiment. All chemicals were obtained from Sigma Aldrich, USA. Confinement solution was identical to the liquid nutrient solution except there was no carbon source. Growth medium and confinement solutions were sterilized in an autoclave at 120 °C in 20 psi pressure for 20 min.

**Channel inoculation by laminar flow-templating**

Following the method of flow templated biofilm growth developed elsewhere, we selectively grew biofilms in either the corners or the middle of the microchannel (Figure 2).[30] This was achieved by exposing the target areas in the microchannel to a so-called "template" flow, with flow rate $Q_T$. Templating solutions included the inoculant solution (1 hour) followed by the AB growth medium (70 h). Template flows were isolated to the target regions of the microchannel with the use of a co-flowing "confinement" flow, with flow rate $Q_C$. Due to the low Reynold's numbers (Re=$7\times10^{-4}$) generated by typical flow rates ($Q_T+Q_C$=1.2 mL·h$^{-1}$) and the small channel dimensions, mass transfer between the two streams was strictly diffusion-limited. In these conditions even co-flowing miscible streams could maintain an interface between them during the 60 s transit time through the microchannel. The choice of the templated growth region was facilitated by injecting the nutrient solution ($Q_T$=0.4 mL·h$^{-1}$) through either inlet I$_1$ (corner growth) or I$_2$ (middle growth) and the confinement solution ($Q_C$=0.8 mL·h$^{-1}$) through the remaining inlet (Figure 1).

**Optical Microscopy**

Optical measurements were acquired using an inverted light microscope (IX73, Olympus, Canada with condenser of numerical aperture, NA=0.9) with fluorescence and white light transmission illumination and 2× magnifications (numerical aperture, NA=0.06). Images were captured using a monochrome CCD camera (Lumenera Infinity 3–1, Ottawa, Canada). Fluorescence images of GFP expressing bacteria were obtained from the same instrument with the use of an appropriate filter cube. Quantitative measurements of the total amount of biofilm were obtained by optical density (OD) from standard microscopy in a defined region of interest within the microchannel. This was based on average pixel intensities at time t ($I_{pixel,t}$) and the average background pixel intensity at t=0 ($I_{pixel,0}$) using the equation for optical density, OD= $-\log(I_{pixel,t}/I_{pixel,0})$.[31] Structural heterogeneity was quantified by coefficient of variance CV=$OD_{STD}$/OD×100%, where $OD_{STD}$ was the standard deviation in the OD within the biofilm-containing portions in the same field of view. To obtain imaging at the attachment surface without interference from out of plain biofilm segments, confocal laser scanning microscopy (CLSM) was conducted (FV1200, Olympus, Canada) with 40× objective (numerical aperture, NA=0.95) and a spectral emission analyzer. Spatial resolution in the z-direction was sub-micron. Excitation of the GFP was at 488 nm and emitted light was detected between 500 and 600 nm with 12-bit intensity resolution. A focus drift compensation unit was used to maintain the same focal plane over long experimental times. In all cases images were analyzed using the Fiji bundle for ImageJ.

**Electrochemical Impedance Spectroscopy**

Electrochemical impedance spectroscopy (EIS) was used to report on the complex electrical response of an electrode-adhered biofilm to imposed oscillations in the applied voltage. Experiments were recorded using a potentiostat (Solartron 1287 Electrochemical Interface, Solartron Electronic Group, UK) in combination with a frequency analyser (Solartron 1260 Frequency Response Analyzer, Solartron Electronic Group, UK) with instrument control software (Zplot, Scribner Associates, Inc, USA). Data treatment and modeling was conducted via software (Zview, Scribner Associates, Inc, USA). For all experiments, the frequency range used was 0.1 Hz–10 kHz. The AC amplitude was 25 mV around a bias potential of 0 V versus Au. The difference in potential between Au pseudo-REs and an Ag/AgCl, KCl (3M) reference electrode was tested 5 separate times by cyclic voltammetry of ferricyanide-containing nutrient solution over a month. The difference in potential remained constant within detectability of the instrument (+/- 0.5 mV) at 360 mV. Separate experiments verified that this difference in potential remained constant as long as the chemical environment did not change, particularly the pH. As the growing biofilm on the WE was downstream of the RE, biofilm related changes to chemical environment near the RE were expected to be minimal (upstream diffusion was calculated to be many orders of magnitude lower than downstream convection). See supporting information for a discussion about the possible effects of RE exposure to biofilm growth in some experiments. All experiments were conducted with the same WE and CE graphite electrodes, which were cleaned with 1M HCl between experiments before re-use. A new RE was fabricated for each experiment.

**Attenuated total reflection infrared spectroscopy (ATR-IR)**

Infrared spectra were recorded using a Nicolet Magna 850 Fourier transform infrared spectrometer (Thermo Scientific, Madison, WI) with a liquid nitrogen cooled narrow-band MCT detector. In order to acquire spectra in attenuated total reflection (ATR) mode, a diamond ATR accessory was used (Golden Gate, Specac Ltd, London, UK). The probing surface of the diamond internal reflection element was 2×2 mm and is fixed in a tungsten carbide holder on the ATR accessory. The resulting spectra were treated using the GRAMS/AI 8.0 software (Thermo Galactic, Salem, NH). The intensity of each band of interest was measured by determining the maximum between a local linear baseline. See Supporting Information for more details.

**Results and discussion**

**Flow templated biofilm growth**

Laminar flow in microchannels enabled the application of steady, well-developed shear forces against wall-adhered biofilms. However, in rectangular cross-section microchannels, wall shear force can change drastically at the intersection between two walls (the microchannel corner), whereas it was nearly constant in the middle portion.[30] Our first goal was to identify how or if these differences in hydrodynamic conditions would affect EIS measurements during biofilm growth. This was conducted by templating biofilm growth against the WE surface in either the microchannel middle or the corners (Figure 2), as explained in the experimental section. The flow rate ratio between confinement and template flows was always $Q_C/Q_T=2$, even for experiments where the total flow rate, $Q_{Tot}=Q_C+Q_T$ was changed. For typical flow rate $Q_{Tot}=1.2$ mL·s$^{-1}$, we estimated the shear stress in the middle portion of the channel to be $\tau = 10.7$ mPa. This was based on the calculated shear stress in a rectangular channel:

$$\tau = 6 \cdot \mu \cdot Q/(w \cdot h^2) \qquad (1)$$

where $\mu$ is the liquid-phase viscosity ($1 \times 10^{-3}$ Pa·s), Q is the total flow rate, and w and h are the channel width (2 mm) and height (300 μm), respectively.

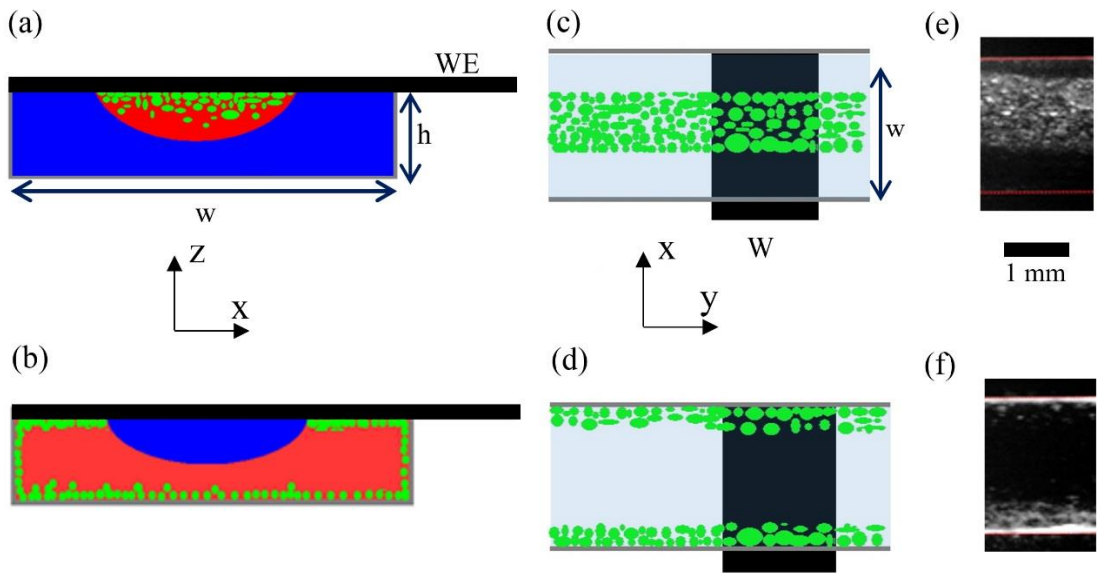

**Figure 2.** Schematic of biofilm growth (green) in the microchannel with width (w) and height (h) under middle (a and c) and corner (d and f) templating approaches. (a) Growth templating flow (red) confined to middle of WE (black) by a confinement flow (blue). The opposite configuration resulted in templating flow confinement to the WE within the microchannel corners (b). Bird's eye view of the middle (c) and corner (d) confined biofilms against the embedded WE. Fluorescence microscopy images of a 27 h old biofilm in the middle (e) and corner (f) of the microchannel with red lines superimposed to show the position of the side walls. Brightness was increased by 25% for visualization purposes. Figures (a) and (b) are in the y,z plane, whereas the other are in the x,y plane. By convention the y axis is in the direction of flow.

Optical microscopy was conducted in parallel with EIS to obtain complementary information on biofilm bulk properties. Growth in the corners and middle of the channel was first registered after 15 h and 24 h and continued until 30 and 50 h, respectively (Figure 3a). Structural heterogeneity continuously diminished in time for biofilms in corners and middle portions of the microchannel until maturity was reached (Figure 3b). After stabilization, structural heterogeneity was about 10 times greater for middle-templated biofilms. We associate this with the fact that biofilms grew thicker in the corners, which resulted in the merging of biofilm segments. In contrast, they remained isolated in the thinner films in the middle of the channels. As microscopy probed the entire vertical cross-section of the biofilm, details specific to the bio-interface were not revealed.

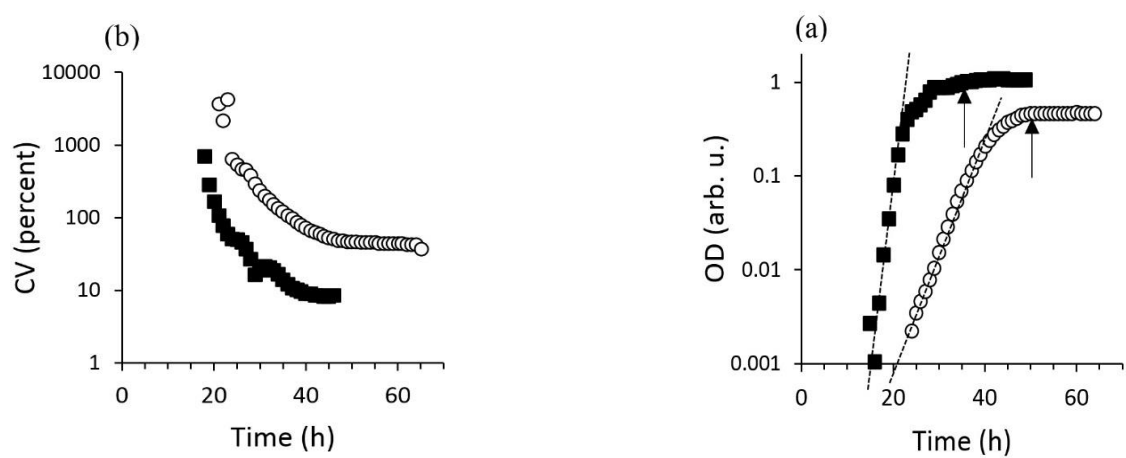

**Figure 3.** (a) *Pseudomonas sp.* growth kinetics visualized by a semi-log plot of OD versus time in channel corners (■) and middle (O). Trend lines show exponential growth period. (b) Structural heterogeneity by coefficient of variance (CV) versus time for corners (■) and middle (O). Arrows point to the curve when the corresponding OD curves were nearly constant. All biofilms were *Pseudomonas sp.* grown with template solution consisting of AB medium containing 10 mM sodium citrate in devices with same dimensions and flow rates.

**Electrochemical impedance spectroscopy on templated biofilms and verification**

In order to study the biofilm directly at the bio-interface, EIS measurements were made in parallel with the optical measurements described above. The interaction between electric fields and charged molecular groups in the biofilm are known to affect the biofilm morphology.[32] Therefore, a single DC offset voltage of 360 mV versus Ag/AgCl was used (see Supporting Information for more details). At this offset voltage, we expected a slight net repulsion of Gram-negative bacteria from the electrode surface.[32, 33] Measurements started 10 h after inoculation and were repeated at time intervals not exceeding 1 $h^{-1}$. Figure 4 shows selected Nyquist plots collected for corner and middle templated biofilms during two separate experiments. In total, Nyquist plots were acquired during four separate experiments conducted at different times for each biofilm templating growth configuration (corner or middle). While biofilms were observed to grow on the CE during the experiment, similar to other EIS studies,[24,26] there should be no effect on the measurements. See Supporting Information for a discussion about the possible effects of RE exposure to biofilm growth in corner templated experiments.

To extract physical meaning from Nyquist curves, we used an equivalence circuit (Figure 4). This circuit was chosen due to its successful application to non-electroactive *Escherichia coli* biofilms in a Luria–Bertani (LB) growth medium in a previous report.[24] Elements included a solution-phase resistor ($R_s$) in series with a parallel circuit. The first branch of the parallel circuit included a resistor, $R_b$, in series with a capacitor, $C_b$, representing bioresistance and bio capacitance of the attached biofilm, respectively. These were the elements used to track biofilm development. On the second branch, elements for solution-phase contributions near the electrode were used. These included a parallel circuit consisting of a charge transfer resistance element ($R_{CT}$) and a constant phase element (CPE), in series with an anomalous diffusion element, $Z_d$. See Section 2 of the SI for more information.

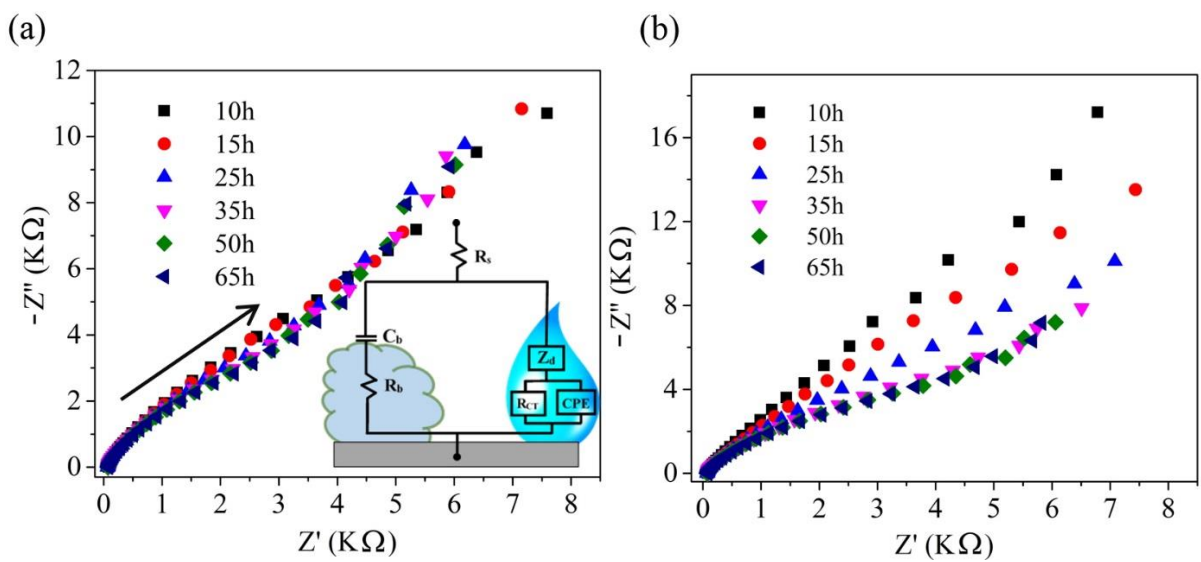

**Figure 4.** Representative measurements of time series Nyquist curves of biofilm growth in the middle (a) and the corner (b) of channel for different times after inoculation. Real and imaginary components of impedance are Z' and Z", respectively. Arrow in (a) points toward lower frequencies. Inset figure in (a) is the equivalence circuit schematic for a non-electroactive electrode-adhered biofilm used in this study. The biofilm portion (left branch) includes biocapacitance and bioresistance ($C_b$, $R_b$, respectively). Liquid-phase interactions at the electrode surface (right branch) include constant phase element (CPE), anomalous diffusion ($Z_d$) and charge transfer resistance ($R_{CT}$). Bulk solution resistance is represented by $R_s$.

We fit the Nyquist plots using the selected equivalence circuit to obtain time-varying values for all fitted electrical components. Figure 5 shows trends for $C_b$ and $R_b$, which relate to the biofilm. We noted the continuously decreasing $R_b$ for both corner- and middle-templated biofilms for the entire experiment. This somewhat counter intuitive result has been observed previously in the first 24 hours after inoculation of an *E. coli* biofilm[24] and was attributed to a growing number of isolated colonies at the electrode surface. In electrical terms, this was considered as a circuit of parallel resistance elements (Figure S2, Supporting Information), which should reduce $R_b$ according to $1/R_b=1/R_1+1/R_2+...1/R_n$, as n increases. The observed increases to $R_b$ after 24 h in the cited study were likely the result of a reduction in n due to merging of enlarged colonies. In the current study, a similar parallel resistor circuit is expected based on widely reported mushroom-shaped biofilms formed by *Pseudomonas sp.* in citrate-containing AB media.[34-37] To account for continuous decreases to $R_b$ observed in this study, we note that n must continuously increase. This can be explained by three mechanisms. First, under similar experimental conditions in microchannels, it has been shown that *Pseudomonas sp.* biofilms continuously emit planktonic bacteria, which can seed uncolonized areas.[35] Second, coulombic repulsion between the WE and the biofilm at the applied potential used here should reinforce localized attachment structures by limiting biofilm/surface contact elsewhere. Third, displacement of the nutrient solution out of the original templated growth region due to accumulation of biofilm, causing templating area to become enlarged in time, further increased n (Fig S3, Supporting Information).

The second tracked bioelectrical element was $C_b$. This is generally considered to be related to the amount of biofilm at the bio-interface. Biofilm accumulation, inferred from increases of $C_b$, was observed immediately after measurements started at 10 h for both corner- and middle-templated growth (Fig. 5b). This shows that EIS was more sensitive to early stages of biofilm accumulation compared to regular optical microscopy experiments conducted with the zoom and numerical aperture used here. The measured $C_b$ continued to increase until 25 h and 20 h for corner- and middle-templated growth, respectively. Following this, the amount of biofilm decreased noticeably for corner templated biofilms, despite growing thicker in these regions. In order to verify that changes to $C_b$ correlated to biofilm surface coverage, we conducted ATR-IR under similar conditions to the EIS experiments (see Supporting Information for more details). Results for the absorbance of the biofilm Amide II band demonstrated growth and a reduction in intensity over nearly the same time intervals as $C_b$ (Figure 5c). Former ATR-IR experiments on *Pseudomonas aeruginosa PAO1* bacteria also demonstrated similar behaviour.[38,39] The reduction in intensity after initial growth was previously attributed to a restructuring event at the attachment surface, which was caused by the migration of bacteria away from surface, likely to maintain closer proximity to nutrient solution.[39] This is supported by CLSM measurements, which confirmed the preference for some bacteria to concentrate near the biofilm/liquid interface, independent of turbulent or laminar flow regimes.[40] For further validation, we conducted CLSM experiments to show that, after an initial growth period, a reorganization process occurred which resulted in an interconnected network of localized biofilm colonies (Figure 5d). See Supporting Information for more information about the time-varying trends to quantified measurements of structural heterogeneity from continuous CLSM imaging. In contrast to surface sensitive techniques (EIS, ATR-IR and CLSM), regular microscopy revealed growth in overall biofilm OD until 30 h for corner growth and 50 h for middle growth (Figure 3). Therefore, the local biofilm accumulation at the electrode surface, and even subsequent restructuring, were both complete while the bulk accumulation of biofilm was still underway.

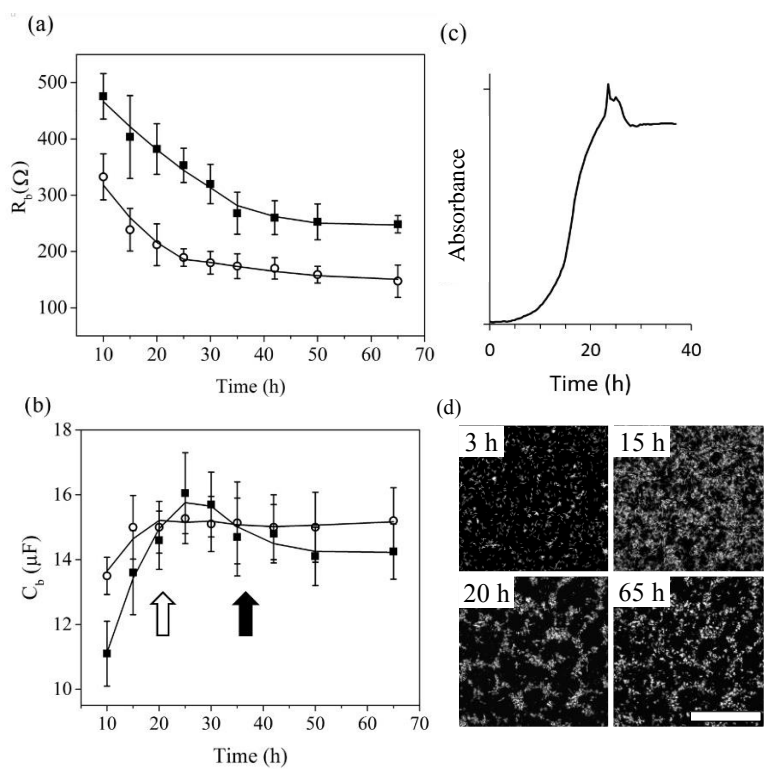

**Figure 5.** Trends of $R_b$ (a) and $C_b$ (b) during templated biofilm growth at the working electrode corner (■) and middle (O). Error bars were produced from the standard deviation of four measurements for each time point. $Q_C/Q_T=0.5$ and $Q_C+Q_T=1.2$ mL·h$^{-1}$ for all measurements. Arrows mark the times from Figure 3 when exponential growth ends for corner (solid) and middle (hollow). (c) Normalized absorption signal from Amide II band in time for *Pseudomonas sp*. Visual confirmation of biofilm restructuring at the attachment surface in the channel middle by CLSM (d) after 3 h, 15 h, 20 h, and 65h using 40× objective. Scale bar (100 μm) in for image at 65 h is representative for all images.

**Biofilm resistance and capacitance under changing flow conditions**

In order to understand the effect of changing flow conditions on surface-adhered biofilm segments, we compared $R_b$ and $C_b$ to optical microscopy results. Middle-templated growth was used in all experiments due to the higher and more uniform shear stress in that part of the channel, as well because no interfering effects could possibly occur from biofilm growth on the RE which was protected by a continuous flow of confinement solution at all times.[30] After 65 h of growth, the biofilm was subjected to different total flow rates in the range 0.6 mL·h$^{-1}$<$Q_{Tot}$<6 mL·h$^{-1}$, corresponding to shear stresses in the range 5.35 mPa<τ< 53.5 mPa (Eq. 1). Changes to $Q_{Tot}$ preserved the flow rate ratio ($Q_C/Q_T$=2), ensuring that the templating dimensions remained constant.[30] Immediately following each EIS measurement, the $Q_{Tot}$ was increased and the system was left to stabilize for 10 minutes until the next measurement. Average values for $R_b$ and $C_b$ were obtained during three different experiments, conducted on separate samples prepared from separate inocula. Both $R_b$ and $C_b$ increased with $Q_{Tot}$ (Figure 6). At the highest flow rate ($Q_{Tot}$=6 mL·h$^{-1}$), $R_b$ and $C_b$ had increased by about 10% over measured values at Q=0.6 mL·h$^{-1}$. Upon returning to the initial $Q_{Tot}$, no statistically significant hysteresis was measured in the $R_b$ or $C_b$ values. We confirmed that the fitted values of liquid-phase equivalence electrical elements remained constant during the shear stress cycling. This indicates that the fluctuation to $R_b$ and $C_b$ were related to changes in the properties of the biofilm directly, not liquid-phase interactions with the electrode surface. To better understand the connection between the biofilm microstructure at the attachment surface and the results obtained by EIS during flow rate cycling, CLSM measurements were conducted on biofilms prepared in exactly the same way as above. After 65 h following inoculation, we observed similar structured colonies as described above (Figure 5d). Next, flow rates were cycled in the same range as above (0.6 mL·h$^{-1}$<$Q_{Tot}$<6 mL·h$^{-1}$) for the same time intervals and CLSM images were acquired at the attachment surface before changing to the next $Q_{Tot}$ value. The results (Supporting Information) show that even at high values of $Q_{Tot}$ the colonies of GFP bacteria maintained their structure, but they shifted by about 20 μm downstream under the applied shear force. After returning to lowest value of $Q_{Tot}$ the bacteria returned to nearly their original position. A slight difference in the initial and final positions (2-4 μm +/- 2 μm) may have been because of the application of elevated sheer forces for times exceeding the biofilm elastic relaxation time.[12] The fact that this displacement was largely reversible indicates that, despite the observed movement of the GFP bacteria, the anchoring point stays fixed upstream. Therefore, we conclude that anchor is made from (non-fluorescent) extracellular polymer material (EPM), which is in line with the generally accepted model of biofilm attachment.[41] We believe that this resulted in temporary structural deformations such as biofilm stretching, thinning and/or modulation of distance between adjacent biofilm colonies, which could cause the observed reversible increases to $R_b$. Moreover, the observed displacement from the upstream anchoring point at elevated $Q_{Tot}$ values indicates a temporary spreading along the attachment surface. This temporary increase to biofilm contact with the substrate supports our observation that $C_b$ increases under elevated flow rates, but returns to original values after flow rate is reduced.

We compared EIS results to micrographs acquired from standard optical microscopy under the same conditions (Figure 6c). After flow rate had increased to $Q_3$ = 2.4 mL·h$^{-1}$, small differences were evident in the images relative to the background. These differences were the result of shifting biofilm density downstream, causing increases to OD on the downstream and decreases to OD on the upstream side of some features. At first this effect was slight and reversible, probably being elastic in nature. However, at $Q_4=4.8$ mL·h$^{-1}$, large and irreversible deformation of biofilm structure was observed in the form of streamers and changes to the surface texture. Streamer formation was attributed to a sudden partial detachment mechanism,[42] whereby portions of locally tall biofilm colonies became partially detached and dangled down-stream. Changes to biofilm texture were also observed and were likely related to the induced viscous flow in the biofilm due to the application of elevated shear forces for times exceeding the biofilm's elastic relaxation time.[12] Therefore, while biofilms at the attachment surface demonstrated reversible changes under elevated bulk flow conditions, biofilm layers further from the attachment surface were more susceptible to irreversible changes.

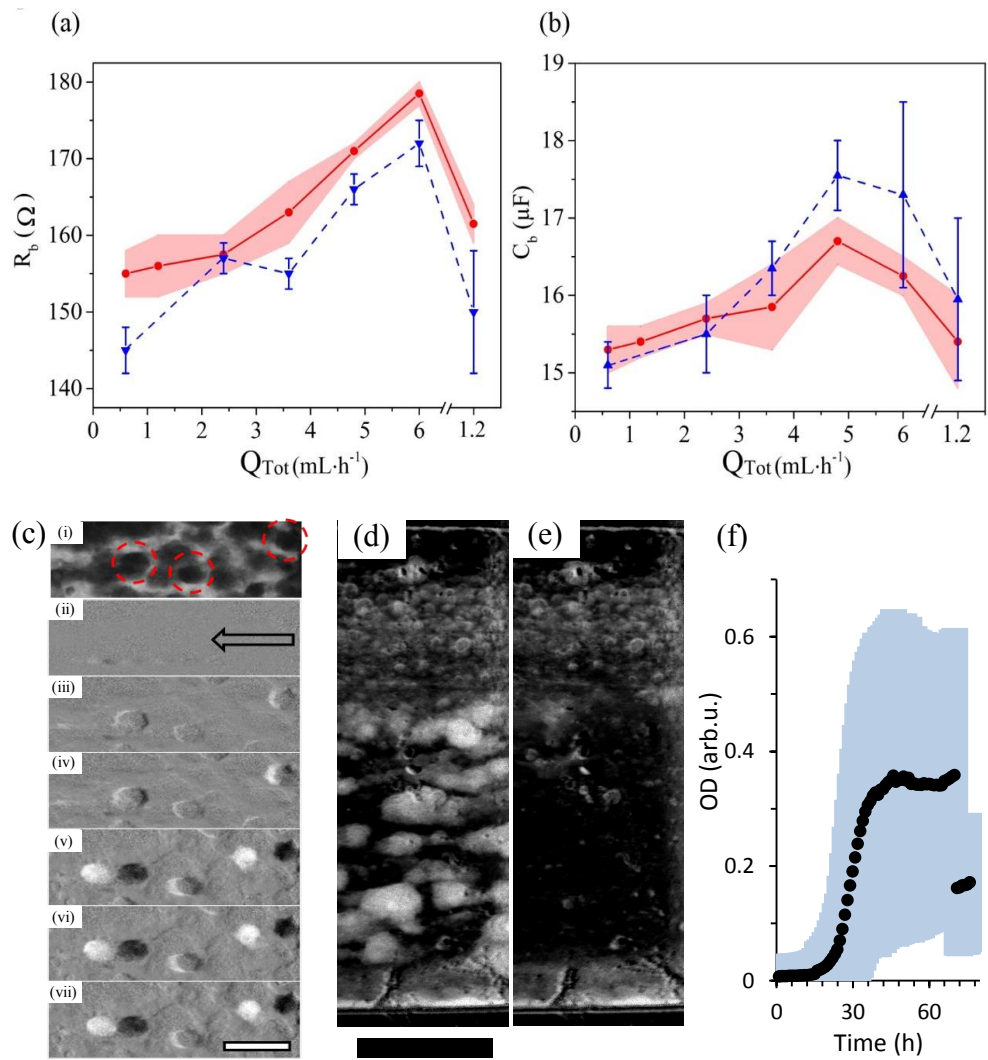

**Figure 6.** Flow rate dependency of fitted bioresistance, $R_b$, (a) and biocapacitance, $C_b$, (b) for 65 h old *Pseudomonas sp.* biofilm that had been cultured under $Q_{Tot}$ =1.2 mL·h$^{-1}$ with flow rate ratio $Q_C/Q_T$=2. Red data points show the average values from three separate experiments, with error bands representing the standard deviation in their measurements. Blue triangles show the trend of resistance and capacitance changes with the same flow rates after shear-removal of significant portions of biofilm. Error bars show standard deviation in their measurements. The last point in each plot shows the $R_b$ and $C_b$ values after returning the system cultivation flow rate of 1.2 mL·h$^{-1}$. (c) Optical micrographs showing typical long-range effects on biofilm due to changing shear

forces. (i) Raw image in region of interest obtained at total flow rate $Q_{tot}$=1.2 mL·h$^{-1}$. Dashed red circles highlight the optically dense biofilm formations, which produced streamer formation under elevated flow in subsequent images. Background corrected images using (i) for the same region of interest at total flow (ii) $Q_{tot}$=0.6 mL·h$^{-1}$, (iii) $Q_{tot}$=2.4 mL·h$^{-1}$, (iv) $Q_{tot}$=3.6 mL·h$^{-1}$, (v) $Q_{tot}$=4.8 mL·h$^{-1}$, (vi) $Q_{tot}$=6 mL·h$^{-1}$, and (vii) after returning to $Q_{tot}$=1.2 mL·h$^{-1}$. In all images, flow is from right to left as indicated by the arrow in (ii). A representative scale bar in (vii) is 250 µm. For all data in (a-c) measurements were obtained 10 min after $Q_{Tot}$ was changed. Optical density mode images of biofilm growth at the electrode-containing microchannel surface before (d) and after (e) biofilm removal via application of high shear stresses. Pixel intensities were calculated from OD=-log($I_t/I_0$), where $I_t$ was the time varying intensity from a specific pixel and $I_0$ is the intensity of the same pixel in the background image. Scale bar is 500 µm. Flow was from right to left. (f) OD vs. time plot of the biofilm growing in the electrode-containing channel during EIS measurements. Error band was obtained from the standard deviation of the pixel intensity from the 2000×610 µm microscope field of view shown in (d) and (e) at each time.

The results above indicate that biofilm layers at the attachment surface were more resilient to shear flow, either because outer biofilm layers provided protection or because the inner layers were naturally more resilient, or both. To directly investigate the effect of prolonged application of shear force directly to surface-attached biofilm layers, we removed large portions of intervening layers and repeated the experiment. Biofilm removal was accomplished by applying a strong pulsing shear force by rapidly modulating the volumetric flow rate of syringe pump between Q=0 and 30 mL·h$^{-1}$ for 10 s. The result was the removal of 50% biomass, mostly from the centre of the channel where the shear stress was the highest (Figure 6e). Following biofilm removal, we observed that $C_b$ values at all $Q_{Tot}$ were statistically similar to measurements before removal (Figure 6b). This indicated that the removal of outer biofilm layers did not affect the quantity of biofilm directly at the electrode surface. Also, the final $C_b$ value after shear cycling against the exposed surface-adhered layers was not statistically different than after the cycling before their exposure. This indicated that the quantity of the surface-attached biofilm was also not affected by direct exposure to range and duration of elevated shear stresses for used here. This is not surprising because carbon electrodes are rough at the microscale, which enables strong attachment of biofilms.[43,44] In contrast, the average values for $R_b$ were all reduced by about 5% compared to those measured before shear removal of outer biofilm segments. Considering that $C_b$ measurements imply that the quantity of biofilm (and therefore the number of colonies) was not changed, the differences in $R_b$ were likely due to changes in biofilm structure following shear removal. We also note that after cycling the shear stress directly against the exposed biofilm seed layers, $R_b$ was reduced by 10% (Figure 6a). This indicates that the structure of the exposed biofilm at the attachment surface was further modified by high flow rates. The nature of these changes should be investigated further, but likely involve the merging of formerly isolated attachment points under applied shear forces, thereby reducing the number of separate colonies.

**Conclusion**

A three-electrode microfluidic flow cell was designed for *in situ* EIS measurements of *Pseudomonas sp.* Measurements of biofilms confined to the channel corners or middle were accomplished by flow templated growth. With the use of an accepted electrical equivalence circuit for non-electroactive biofilms, bioresistance and biocapacitance were measured in time.

Using EIS and verification by ATR-IR and CLSM, biomass accumulation and a restructuring process at the surface were finished within the first 25 hours, whereas regular optical microscopy only showed continuous growth in the biofilm bulk until much later. Finally, we applied different shear stresses against middle-templated biofilms by cycling the total flow rates. The EIS results showed a strong, reversible correlation to the measured bioresistance and biocapacitance. Supporting measurements by CLSM suggest temporary stretching and spreading of the biofilm were responsible for the EIS observations. Regular optical measurements showed that cycling shear stress caused irreversible changes to bulk biofilm structure. Shear removal of the biofilm outer layers did not affect the amount of biofilm measured by EIS, but changes in bioresistance suggest that its structure at the attachment surface was changed and could be further modified by application of elevated shear stress directly against the exposed surface adhered layers.

AUTHOR INFORMATION

**Corresponding Author**

Jesse Greener, Laval University, Department of Chemistry, Pavillon Alexandre-Vachon, 1045, avenue de la médecine, local 4064$^E$. Québec (Québec). jesse.greener@chm.ulaval.ca

**Author Contributions**

Together, MPZ, JG, SJC conceived of the concepts and planed the experiments. MPZ fabricated flow channels, conducted all electrochemical experiments and analysis and some CLSM experiments and analysis. FPM conducted ATR-FTIR and some CLSM experiments and analyzed them. The manuscript was written by JG with contributions from all authors. All authors have given approval to the final version of the manuscript.

**Funding Sources**

This research was funded by grants to JG from Fonds de Recherche du Québec - Nature et Technologies ("nouveau chercheur" grant), Natural Sciences and Engineering Research Council of Canada (NSERC) Discovery grant and Canadian Foundation for Innovation (CFI) Leaders opportunity fund. SJC is a research scholar from the Fonds de Recherche du Québec en Santé.

ABBREVIATIONS

EIS, electrochemical impedance spectroscopy; PDMS, polydimethylsiloxane; ATR-IR, attenuated total reflection infrared spectroscopy; WE, working electrode; RE, reference electrode; CE, counter electrode; OD, optical density; CV, coefficient of variance;

For Table of Contents Only

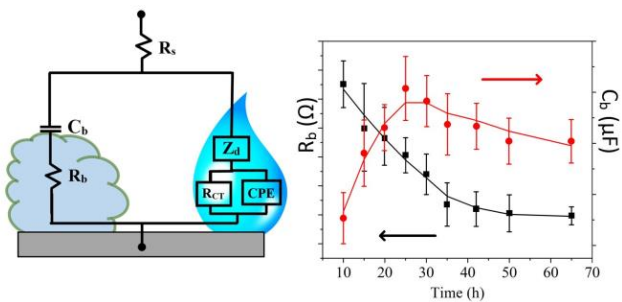

Table of contents graphic.